# Andreev reflection spectroscopy of the new Fe-based superconductor $EuAsFeO_{0.85}F_{0.15}$: evidence for the strong order parameter anisotropy

V.M.Dmitriev[1,2,*], E.P.Khlybov[1,3], D.S.Kondrashov[2], A.V.Terekhov[2], L.F.Rybaltchenko[2], E.V.Khristenko[2], L.A.Ishchenko[2], I.E.Kostyleva[1,3], and A.J.Zaleski[4].

1. International Laboratory for High Magnetic Fields and Low Temperatures, Gajowicka 95, 53-421 Wroclaw, Poland.

2. B.I.Verkin Institute for Low Temperature Physics and Engineering, NAS of Ukraine, 47 Lenin Ave., Kharkov, 61103, Ukraine.

3. L.F.Vereshchagin Institute for High-Pressure Physics, RAS, Troitsk, 142190, Russia.

4. W.Trzebiatowski Institute of Low Temperature and Structure Research, Polish Academy of Sciences, P.O.Box 1410, 50-950, Wroclaw, Poland.

## Abstract

Andreev reflection spectra have been measured in a new superconductor $EuAsFeO_{0.85}F_{0.15}$ having an unexpectedly low superconducting transition temperature $T_c \approx 11.3$ K among related FeAs compounds on a base Sm and Gd surrounding Eu in the series of lanthanides. The nearly fivefold lower $T_c$, as against the expected value, is attributed to the divalent properties of Eu ions when in the compound investigated along with the weakly magnetic $Eu^{3+}$ ions may be present and the strongly magnetic $Eu^{2+}$ ones that is a strong destructive factor for superconductivity. Most of the spectra measured showed features that corresponds to two energy gaps whose values varied from contact to contact within $2\Delta_s/kT_c = 2.2 \div 4.7$ and $2\Delta_l/kT_c = 5.1 \div 11.7$ for small and large gap, respectively. The corresponding variations for single-gap spectra are $2\Delta/kT_c = 2.6 \div 6.4$. The relatively large size of crystallites (no less than ~25 μm) and the large number of contacts measured (several tens) suggest with a high degree of probability that the spectra obtained account quite fully for the gap distribution practically in all crystallographic directions. The data obtained and the absence of zero gaps in the measured spectra evidence in favor of the anisotropic s- or s±-symmetry of the order parameter in $EuAsFeO_{0.85}F_{0.15}$ that was revealed in other similar compounds with higher $T_c$. Thus, the character of the gap function $\Delta(\mathbf{k})$ in this compound is inconsistent with the d-wave superconductivity observed in some low-$T_c$ pnictides.

# Introduction

The discovery of a basically novel high–$T_c$ $LaO_{1-x}FeAsF_x$ superconductor with the onset of the superconducting transition at $T_c \approx 26$ K [1] has stimulated a search for other similar compounds (briefly denoted as 1111–type systems). In some cases, substitution of La with other Ln–series elements (Ln – lanthanide) raised $T_c$ significantly, for example, to $T_c \approx 55$ K for Fe-based 1111 compound with Sm [2] and to $T_c \approx 54$ K for compound with Gd [3]. As the temperature lowers, the parent LnOFeAs compounds consisting of alternating LnO and FeAs layers undergo structural and successive/simultaneous antiferromagnetic (AFM) transitions in the interval 160-180 K. The transitions can be suppressed when O is partially substituted by F. On such substitution excessive electrons appear in the LnO layer, which then pass over to the FeAs layer and activate the superconducting state there.

Later on superconductivity was also observed in other FeAs similar systems which contained no oxygen. Three–component $A$$Fe_2As_2$ (briefly 122) [4] compounds are such systems in which the FeAs layers have practically identical crystalline structures. In these systems superconductivity appears when divalent element $A$ (Ba, Ca, Sr) is partially substituted with a univalent one (usually K) that induces a hole doping of the FeAs layers. The highest $T_c \approx 38$ K is achieved at $A = Ba_{1-x}K_x$. The physical properties of both types of superconductors are quite similar, but the preparation technology of 122 compounds is much simpler. Besides, superconductivity was detected in some materials that do not need doping (e.g., LiFeAs with $T_c \approx 18$ K [5]) and in the non-stoichiometric monolayers of Fe chalcogenides $FeX_{1-x}$ (X=Se, Te) with $T_c \approx 8$ K [6].

The discovery of high-$T_c$ superconductivity in Fe-containing compounds has initiated intensive investigations aimed at clarifying the mechanism of the Cooper pairing and the symmetry of the superconducting order parameter. The nature of electron attraction in these compounds is not yet clear completely and the preference is given mainly to the magnetic mechanism. Most of the experimental results obtained on superconducting Fe pnictides show considerable variations of the gap near the Fermi level, though it never turns zero. This suggests the existence of the anisotropic s-wave type gap function in these compounds. Nevertheless, nodes or lines of nodes of the gap observed in some compounds, when partially substituting Co for Fe or P for As, and in a number of low-$T_c$ pnictides that evidences the d-wave symmetry of Cooper pairing.

Andreev reflection spectroscopy of the N-S point contacts is one of the simplest and sufficiently reliable methods of estimating the value and symmetry of the order parameter (gap) in various kinds of superconductors. It has the advantage of finding the gap structure in different crystallographic directions avoiding intricate mathematical procedures. By such the technique,

single BCS-like gap $2\Delta_0/kT_c \approx 3.7$ in $SmO_{0.9}FeAsF_{0.1}$ was obtained for the first time by Chen et al. [7]. However, most of the subsequent investigations on 1111 systems revealed two gaps, each varying widely for the same compound [8-11]. For example, in $NdO_{0.9}FeAsF_{0.1}$ the small and the large gaps varied within $2\Delta_s/kT_c=1.8\div2.7$ and $2\Delta_l/kT_c=4.1\div5.9$, respectively.

Such a scatter of gaps found by different authors may be due to the anisotropy of the gap function in the **k**-space in different sheets of the Fermi surface (FS). When the number of probes is small (which is for some reasons typical of PC spectroscopy), only a limited number, if not single, of crystallographic directions are scanned. Therefore, the gap values measured in different investigations do not coincide. One more factor – the quality of the sample especially its surface – is no less important. With an improper control over the onset of the superconducting transition in each contact, its central part may contain a region with a disturbed stoichiometry or significant surface contamination. In this case the resulting PC spectra will not display the characteristics of the bulk sample.

In this study the Andreev reflection spectra have been investigated in point contacts based on the polycrystalline $EuAsFeO_{0.85}F_{0.15}$ compound having an unexpectedly low $T_c \approx 11.3$ K, as against other 1111-type systems. The large-size crystallites (no less than ~25 μm) and a great number of measured contacts (several tens) give reasonable confidence that the spectra obtained account quite fully for the gap distribution practically along all crystallographic directions. Both one–gap and two-gap spectra (in most cases) were observed in our $Au-EuAsFeO_{0.85}F_{0.15}$ contacts. In the one-gap spectra the relative gap varied within $2\Delta/kT_c=2.6\div6.4$. In the two-gap spectra the relative small $\Delta_s$ and the relative large $\Delta_l$ gaps varied within $2\Delta_s/kT_c=2.2\div4.7$ and $2\Delta_l/kT_c=5.1\div11.7$, respectively. For any of these contacts the ratio $\Delta_l/\Delta_s$ was within $2\div4$. The results obtained and the absence of zero gaps in the spectra measured evidence in favor of the anisotropic s-wave (or s±-) symmetry of the order parameter in $EuAsFeO_{0.85}F_{0.15}$, which was previously revealed in other similar compounds.

Among the abundance of information about Fe-based oxypnictides, we have failed to find at least one report of synthesizing a Eu–containing 1111 compound. Because of the Eu position in the periodic table of the elements between in Sm and Gd, which are constituents of the 1111 systems with $T_c > 50$ K [2,3], such attempts might be made but possibly with no success. At the same time, there are many publications about using Eu for fabricating the 122 systems with relatively high $T_c > 30$ K. This may be because Eu, like most lanthanides, is a polyvalent metal having 2+ or 3+ valence in different chemical compounds. However, unlike other lanthanides, the lower valence of Eu is preferable for forming metallic bonds, such as in 122 systems, where under certain conditions the divalent metallic layers dope holes to the FeAs layers generating superconductivity in them. In 1111-type compounds doping electrons come to the FeAs layers

from the adjacent lanthanide oxide ones, where Ln should be in trivalent state. Such a Ln-state is typical of the compounds with strongly electronegative metalloids, for example, F or As. This sort of compounds is normally present in the mixture of the starting components for synthesis of 1111 systems.

## Experiment

Such the ingredients as $EuF_3$, $EuAs$, $Fe_2O_3$ and $Fe$ were used for preparing the $EuAsFeO_{0.85}F_{0.15}$ compound. The chemical solid–phase reaction proceeded in an Ar–filled quartz ampoule at $T=1150^0$ C for 24 hours. For homogenization, the samples were ground and kept at this temperature for 30 hours. As was expected, with this technological processing Eu should retain its trivalent state. The typical curve describing the resistive transition to the superconducting state in one of the samples is illustrated in Fig.1. (A similar transition was also registered in the temperature dependence of magnetic susceptibility). Of surprise is the unexpectedly low $T_c \approx 11.3$ K (the onset of the transition) in comparison with other Fe-based 1111 compounds, including the neighboring rare-earth elements Sm and Gd, with $T_c>50$ [2,3]. (Eu is located between Sm and Gd in the lanthanide series).

Previously [12] we tried to correlate the low $T_c$ of $EuAsFeO_{0.85}F_{0.15}$ with the atomic radius of Eu, which is rather large in comparison with other lanthanides. This assumption seems rather doubtful because literature data on the atomic radius of Eu are rather controversial. On the other hand, the literature data on the ionic radius accounting most accurately for ionic bonds are practically identical for the trivalent state of these elements. It is therefore reasonable to assume that decrease in $T_c$ is due to the magnetic, rather than structural factor. The point is that $Eu^{2+}$ ions have the largest spin magnetic moment $S\sim7\mu_B$ among the lanthanide elements. This feature is determined by the fact that a half of the f-electrons (14 altogether) have identical spin orientations precisely corresponding to the Hunde rule. In the trivalent state one electron leaves the f-shell, which decreases the spin moment and induces an orbital moment which partially compensates for the spin one. We are unaware of the total magnetic moment in $Eu^{3+}$ ions but, according to the indirect evidence, it can be high enough. Besides, because of the mixed-valence effect typical of many rare-earth compounds, $EuAsFeO_{0.85}F_{0.15}$ can also contain $Eu^{2+}$ ions, which enhances the destructive influence of magnetism on singlet superconductivity. It is quite possible that the presence of $Eu^{2+}$ ions provides a certain level of hole doping which counterbalances the main mechanism - electron doping and thus decreases the effective number of carriers in the FeAs layer, hence $T_c$ as well.

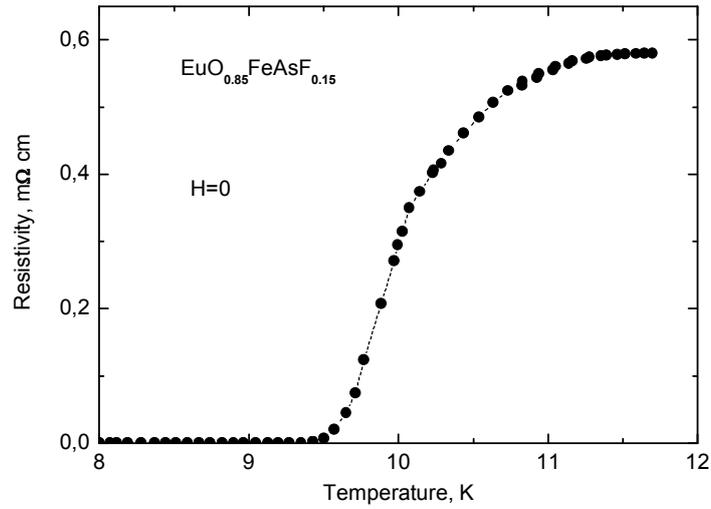

Fig.1. Resistive transition to the superconducting state in the EuAsFeO$_{0.85}$F$_{0.15}$ compound investigated in this study.

The Andreev reflection spectra, dI/dV(V) characteristics, were measured on point contacts (PC) having metallic conductivity (without an additional insulating interlayer) between a mechanically–sharpened chemically-polished Au needle (N-electrode) and freshly–fractured surface of EuAsFeO$_{0.85}$F$_{0.15}$ (S-electrode). The S-electrode consisted of small (2-3mm across) pieces broken off from a sintered bulk. The fracture was a conglomeration of brilliant crystallites about 100 μm across. Some of them were split into smaller (~25 μm) blocks with small-angle misorientation. Besides, the fracture had dull areas possibly of amorphous slag that took about one-third of the fracture surface. As a result, the share of the superconducting phase could hardly exceed two thirds of the sample volume. Taking into consideration the 100% superconducting screening, we can state that the dull areas do not degrade the electric contact between individual crystallites. The comparatively small width of the superconducting transition rules out significant variation of the superconducting parameters over the sample volume.

The electrodes were brought together in liquid He. A special device was used to move the electrodes relative to each other in two perpendicular directions. We were thus able to change the point of contact on the S-electrode without heating the sample. The PC spectra were registered using the standard modulation technique of lock-in detection at the frequency 437 Hz. With this technique we could make contacts in a wide interval of resistance. To preclude thermal effects and to ensure good mechanical stability, the preference was given to point contacts with moderate resistance scatter (2-10 Ω). Most of such contacts exhibited a spectroscopic regime,

which is proved by high–level excess (Andreev) current close in some cases to the theoretical value, which did not change up to the voltage no less than, at least, several $\Delta/e$. Some spectra had additional features at $eV \gg \Delta$, which were most likely due to the reduced critical current in the inter-crystallite layers typically observed in materials prepared by solid-state synthesis and could not influence the basic (informative) portion of the spectra.

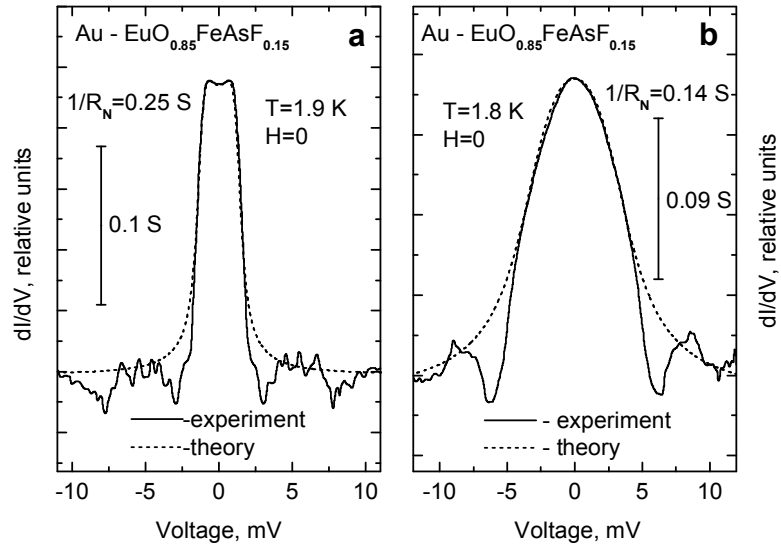

Fig.2. Two typical one-gap dI/dV(V) spectra of Au-EuAsFeO$_{0.85}$F$_{0.15}$ point contacts (solid lines - experiment, dash lines - BTK fitting) differing in gap size and degree of smearing of the spectral lines ( fitting parameter $\Gamma$). a): $1/R_N \approx 0.25$ S, $\Delta \approx 1.3$ meV, $\Gamma \approx 0.01$ meV, $Z \approx 0.1$;
b): $1/R_N \approx 0.14$ S, $\Delta \approx 3.1$ meV, $\Gamma \approx 1.8$ meV, $Z \approx 0.15$.

The size of the contact can be found by using the Sharvin formula which corresponds to the ballistic regime. However, this is impossible in our case because the Fermi parameters are not known for the compound investigated. We estimated no more than the upper limit of contact sizes using the Maxwell formula $d=\rho_0/R_N$ most suitable for diffusive regime. It is obvious that the residual resistivity $\rho_0$ of the bulk sample (Fig.1) is excessively large. Most likely this is due to the influence of the aforementioned intercrystalline layers having poor electric conductivity. Therefore, this $\rho_0$ does not account for the electric properties of the crystallites themselves.

The calculation of d using $\rho_0$ of Fig.1 would yield anomalously large micron-scale sizes of the contact, which is not compatible with the spectroscopic character of the registered spectra. The real sizes of the contact can be obtained from $\rho_0$ measured on single crystals whose properties may be similar to those of the crystallites of the sintered samples. Unfortunately, this kind of information is unavailable for EuAsFeO$_{0.85}$F$_{0.15}$. As for the electric characteristics of

single crystals of similar pnictides, their residual resistivity is known to be over one and a half order of magnitude lower than that presented in Fig.1. For example, for single crystalline LaFePO $\rho_0$ is about 5 $\mu\Omega$cm [13]. In this case the calculation of the contact sizes would give quite reasonable values within approximately 5÷25 nm. This is a rough estimation of the upper limit of our contact sizes because the Maxwell formula yields essentially larger values than the Sharvin one.

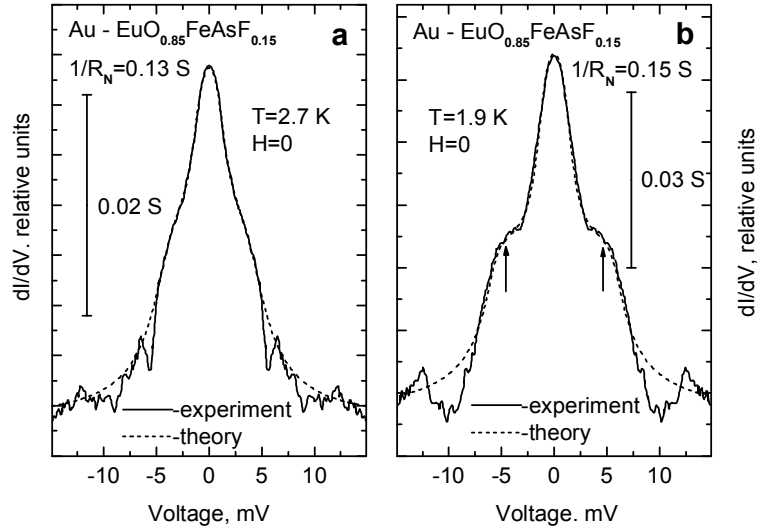

Fig.3. Two examples of dI/dV(V) spectra that can be fitted accurately in the two–gap BTK model [25]: a): $1/R_N \approx 0.13$ S, $\Delta_s \approx 1.1$ meV, $\Gamma_s \approx 0.4$ meV, $\Delta_l \approx 4.3$ meV, $\Gamma_l \approx 1.1$ meV, $Z \approx 0.1$;
b): $1/R_N \approx 0.15$ S, $\Delta_s \approx 1.3$ meV, $\Delta_l \approx 5.9$ meV, $\Gamma \approx 0.5$ meV, $Z \approx 0.1$.

The order parameter $\Delta$ was estimated on the basis of the Blonder–Tinkham–Klapwijk (BTK) theory [14] which provides an adequate description of the electric characteristics of N-S contacts. The experimental PC spectra were fitted to the modified BTK formulae [15] including an additional parameter $\Gamma$ characterizing the Cooper pair lifetime [16], which defines the smearing of the spectra in the region of gap energies. In reality this parameter also accounts for the effects of the crystal structure imperfection in the contact area which can cause an inhomogeneous distribution of the order parameter at submicron-scale dimensions. There is also another parameter Z characterizing a possible potential barrier at the N-S interface that can be generated by the dielectric interlayer or by the discrepancy between the Fermi parameters on both sides of contact.

## Results and discussion

All of the measured electric characteristics (spectra) dI/dV(V) of Au-EuAsFeO$_{0.85}$F$_{0.15}$ point contacts had spectral features that pointed to a high Andreev reflection intensity close in many cases to the theoretically predicted value. Some spectra had the standard form typical of traditional one–band superconductors with a single gap. However, in most cases the registered spectra could be described only in the two-gap approximation. Fig.2 illustrates two spectra of the first type whose BTK fitted [15] gap parameters differ considerably: $\Delta_{min} \approx 1.3$ meV (Fig.2a) and $\Delta_{max} \approx 3.1$ meV (Fig.2b). The corresponding characteristic ratios $2\Delta/kT_c$ are 2.6 and 6.4, respectively. (The estimates were obtained for $T_c \approx 11.3$ K, corresponding to the onset of the superconducting transition). These data point to high anisotropy of the order parameter in EuAsFeO$_{0.85}$F$_{0.15}$.

Of interest is the low intensity of the double maxima in the gap energy region (Fig.2a) or even their absence (Fig.2b). It is known that such maxima are always observed in the spectra of N-S contacts based on traditional s-wave superconductors when the Fermi velocities are different in the two electrodes and/or there is a thin dielectric interlayer in the contact area (Z>0), as is stated in the BTK theory. Assuming that the Fermi velocity of the electrons is low in Fe oxypnictides, we could expect the intensive double maxima or even a tunnel regime (Z>>0) in our contacts. This has not occurred. Note that low intensity of this structure was also observed in other Fe arsenides [17-19].

This discrepancy between theory and experiment was also observed in contacts based on the superconducting copper-oxide and heavy-fermion compounds. The phenomenon was analyzed by Deutscher and Nozieres [20] who assume that the electron mass renormalization responsible for the effective Fermi velocity is much weaker in the PC region than in the bulk material. A detailed analysis of the processes of quasiparticle transition and relaxation in the contact area using Green function technique supported the assumption at the microscopic level. The original BTK theory contains some simplifying assumptions which disregard the real distribution of the pairing potential at the N-S interface and the electron structure of the superconductor. In this context it is hardly possible to calculate correctly the effective Fermi velocity in multiband superconductors using the parameter Z from the BTK analysis of PC spectra.

This is evident in the calculation of the Fermi velocity $v_{FS}$ of EuAsFeO$_{0.85}$F$_{0.15}$ based on the formula following from the BTK theory [21]

$$Z^2 = [Z_0^2 + (1-r)^2/4r]$$

where $r = (v_{FS}/v_{FN})$. The estimation was made using the barrier parameter $Z=0.1-0.15$ for the contacts in Fig.2. Since the Andreev current is high, the possible dielectric layer at the N-S boundary can be neglected and hence $Z_0$ can be assigned zero. Taking $v_{FN}=1.4*10^8$ cm/sec for Au, we obtain only a ~20-30% decrease in $v_{FS}$, which is unlikely for iron pnictides. According to the data published for some compounds of this family, $v_{FS}$ varies within ~(0.3-2.4)*$10^7$ cm/sec in different sheets of the Fermi surface [22,23]. The significant (2- to 9-fold, according to different sources) increase in the free electron mass, calculated from experimental data on the photoemission spectroscopy, de Haas-van Alphen effect and heat capacity [24,25] for different crystallographic directions, is another point in favor of low $v_{FS}$.

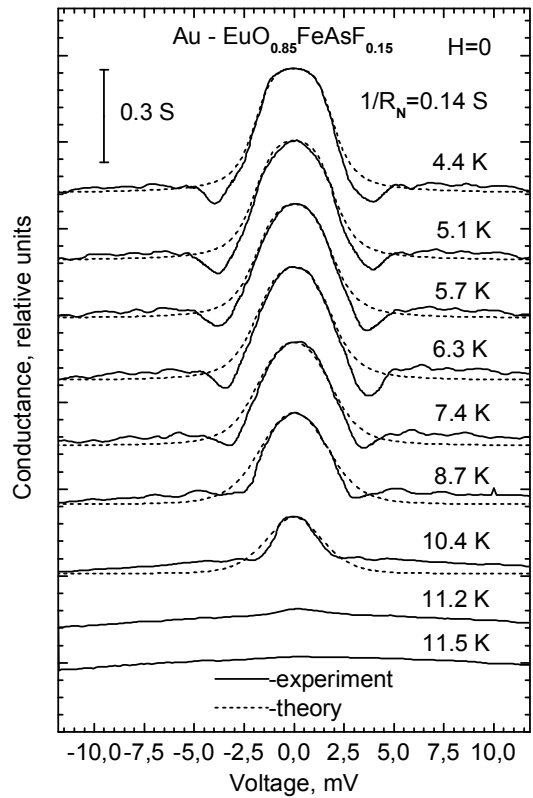

Fig.4. A temperature set of dI/dV(V) spectra taken on a one–gap contact; $\Delta(4.4K)\approx1.5$ meV, $\Gamma\approx0.01$ meV, $Z\approx0.1$.

The low electron mass renormalization in the contact region (as follows from the BTK analysis of PC spectra for such classes of nonconventional superconductors as copper oxides, iron pnictides and heavy-fermion systems) may also be dependent on the type of Cooper pairing. Let us assume that Cooper pairs are formed by some other (e.g., magnetic) mechanism different from the phonon one, as is postulated by the classical Bardeen-Cooper-Schrieffer (BCS) theory. In this case magnons along with phonons would participate in the scattering processes involving the superconducting excitations (bogolons). As a result, the relative part of the electron-phonon scattering events could be reduced. We believe that the contribution of the electron-magnon interaction to the electron mass renormalization in iron pnictides cannot be large. Thus, the small height of the potential barrier in the contacts based on multiband superconductors can be attributed both to the specific transition of various types of charge carriers trough the N-S boundary [20] and to a non-phonon mechanism of Cooper pairing.

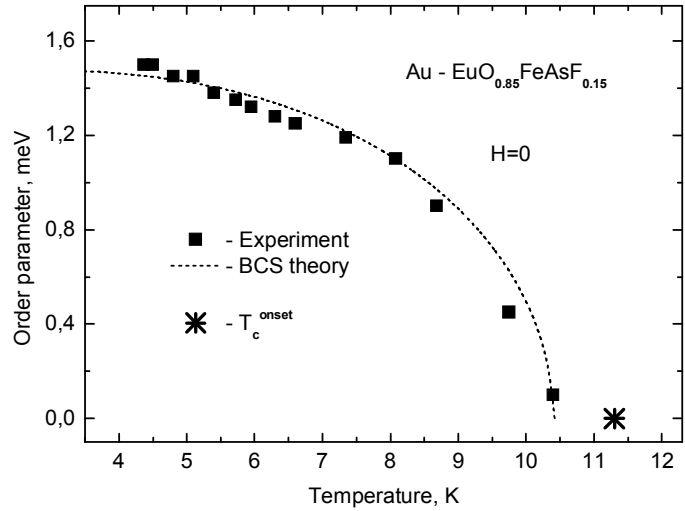

Fig.5. The temperature dependence of the gap parameter $\Delta(T)$ obtained from the BTK fitting of the one–gap spectra of Fig.4.

The structure of most of the measured spectra had additional (as compare to conventional superconductors) features in the region of gap voltages. Therefore, the BTK-fitting of these spectra in the one-gap approximation induces a large error. It is reasonable to relate the additional features to the second energy gap. Two typical spectra with nearly equal contributions of each gap to the excess current are shown in Fig.3 (arrows mark the second gap-related features, Fig.3b).

The possibility of revealing two gaps in a two-band superconductor was demonstrated convincingly in 2001 by Szabo et al. [26] for the first time through measuring the Andreev reflection spectra in $MgB_2$-based N-S contacts. In the line with this study we separated the experimentally observed Andreev reflection amplitude into two components assuming that these parts take contributions from different sheets of the Fermi surface. Each component was then BTK-fitted.

First, the low-energy part of the spectrum was fitted, which enabled us to use the obtained barrier parameter Z for fitting the high-energy part of the spectrum, independent estimation of Z being impossible for the spectra registered. The procedure used is quite reasonable because the barrier height can hardly vary in a very narrow energy interval (several meV). The smearing parameter Γ was not always identical for both the parts of the spectrum and this is quite normal because the intensity of quasiparticle scattering at impurities and structural defects can differ essentially from band to band. And this is not surprising since in the nonconventional superconductors inelastic scattering can initiate the pair-breaking effects, which have been reliably established for oxide high-$T_c$ compounds. Recently, the well justified assumptions about the existence of a similar effect in the iron pnictides have appeared.

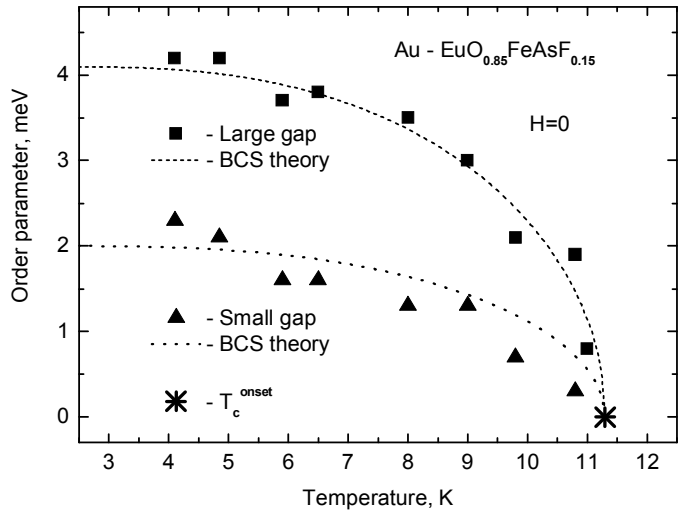

Fig.6. The temperature behavior of the small and large gap parameters specified in the BTK analysis of one of the contacts exhibiting two-gap features; $\Delta_s(4.1K)\approx 2.3$ meV, $\Delta_l(4.1K)\approx 4.2$ meV

The BTK formulas were used to calculate the conductances $\sigma_s(eV)$ and $\sigma_l(eV)$ dependent on the small and large gaps, respectively. At the final stage the calculated total conductance $\sigma=\alpha*\sigma_s+(1-\alpha)*\sigma_l$ was fitted to the experimental one to find the relative contribution of each gap, that is, weight factors $\alpha$ and $(1-\alpha)$. For most contacts these contributions were quite close $\alpha=0.45-0.55$. The estimated small $\Delta_s$ and large $\Delta_l$ gaps for different contacts varied within 1.1÷2.3 meV ($2\Delta_s/kT_c=2.2÷4.7$) and 2.5÷5.9 meV ($2\Delta_l/kT_c=5.1÷11.7$), respectively. (The ratio $\Delta_l/\Delta_s$ for any contact was within 2÷4). It is obvious that the upper bounds of $2\Delta/kT_c$ intervals are inconsistent with the phonon model of Cooper pairing and an alternative mechanism should be considered for the compound investigated.

It is unlikely that the variations of gap parameters are caused by an inhomogeneous distribution of the critical parameters over the sample volume. The processed spectra show that the superconducting transition started at practically the same temperature in all the contacts ($T_c\approx 11.3$ K), implying the invariability of order parameter over a contact area. This also indicates that only one, sufficiently perfect, crystallite is present in the contact area. If the contact spot occurred at the joint of two (or three) neighboring crystallites, the related spectra usually had parasitic features of non-spectroscopic character because of the weakened electric coupling between the crystallites. Such spectra are not considered here. We thus believe that the information about two gaps arrives from the same microscopic volume of the sample.

The considerable variations of the gap parameters observed in the study can be attributed to the anisotropy of the gap function $\Delta(\mathbf{k})$ in $EuAsFeO_{0.85}F_{0.15}$. This idea was supported in a number of experiments on different Fe-based 1111- and 122-type compounds. For example, $\Delta_s$=1.5 meV and $\Delta_l$=9 meV were obtained in μSR measurements on single crystalline $(BaK)Fe_2As_2$ ($T_c$=32 K) [27]. The measurement of the first critical field in a similar compound gave $\Delta_s$=2.0 meV and $\Delta_l$=8.9 meV [28]. The results obtained in the NMR [29] and angle-resolved photoemission [30] experiments as well as in measurements of thermal conductivity [23] also point to the order parameter anisotropy in iron pnictides. Theoretically (e.g., [31-33]) these experimental facts are explained within the anisotropic s±-wave model of Cooper pairing in the Fe-based superconductors.

In PC measurements the relative contribution to the spectrum from individual sheets of the Fermi surface, where superconductivity is realized, can depend on the orientation of the contact axis relative to the crystallographic directions of the probed crystallite. Despite the polycrystalline structure of the samples, each crystallite (subcrystallite), no less than ~25 $\mu$ in size, is actually a small single crystal. The crystallite size exceeding the expected contact diameter allows the regime of directional spectroscopy of the order parameter along some, even if uncertain, crystallographic axis. Such spectroscopy is feasible, at least in a rough approximation, due to the limited width of the bunched beam of quasiparticles incoming to the N-S boundary. The limitation is caused by the large difference between $k_F$ values in both electrodes as well as a contact geometry, which actually represents the narrow elongated channel. Unfortunately, we cannot estimate the angular selectivity of the PC technique used in this study because $k_F$ of the compound investigated and the exact geometry of the constriction formed when the electrodes come into a contact are unknown.

As follows from many photoemission experiments and theoretical calculations of the band structure of iron pnictides, there are two hole FS cylinders at the center of the Brillouin zone (Γ point) and two cylindrical electron FS sheets at the corners of the zone (M points). It is found that the angular distribution of the order parameter is anisotropic at least in the electron sheets.

We believe that the observation of one- or two-gap spectra is dependent on the orientation of the contact axis relative to the principal crystallographic axes. When the contact axis is close to the ΓX-direction, the PC spectrum can exhibit the one-gap features (Fig.2a). The reason is that in this case only hole FS sheets near the Γ-point can form the structure of the spectrum. (Note that there are no other FS sheets near the X-points that are between the M-points in the Brillouin zone). Correspondingly, a two-gap spectrum is expected along the ΓM-direction (Fig. 3b). Such a spectrum can furnish information about the averaged gap values on both the hole FS sheets around the Γ-point and the electron FS sheets near the M-point. In a certain approximation, the spectra in Figs.2a and 3b can be perceived as quite "pure". In these spectra the parameter Γ characterizes not only the spatial distribution of the order parameter typical of anisotropic superconductors but the processes of Cooper pair breaking as well, which determines the broadening of the gap itself. (See Fig..2a, Γ≈0.01 meV and Fig.3b, Γ≈0.5 meV).

For the contacts measured in the intermediate directions (between ΓX- and ΓM-lines) the peripheral areas of the electron FS sheets situated near the M-point can modify significantly the one-gap spectral lines present in all spectra. (The latter are determined only by hole FS sheets located around the Γ-point). The spectra appeared to be smeared appreciably and were characterized by a sharply increased fitting parameter Γ. In this case the parameter accounts essentially for the non-uniform angular distribution of the order parameter and to a much lesser degree for its broadening caused by the breaking of the Cooper pairs. Evidently, the spectra shown in Fig.2b (Γ≈1.8 meV) and Fig.3a (Γ≈1.1 meV) belong to this type.

The considerable scatter of the gap parameters measured in different PC experiments on identical FeAs compounds published so far (see survey [11]) may be connected not only with the quality of the samples, but with anisotropy of the gap function near the Fermi-level as well. The effects of intra- and inter-band scattering of quasiparticles at impurities can also influence the order parameter value, because in nonconventional superconductors the elastic scattering of quasiparticles has pair-breaking effect. Note that none of the measured spectra had close-to-zero gap parameters, which indicates the absence of zeros or lines of zeros in the Δ(**k**)-dependence in $EuAsFeO_{0.85}F_{0.15}$. Thus, the assumed [34,35] existence of zero gaps in some low temperature pnictides lacks a support in this case.

For one-gap spectra the gap was estimated quite accurately at different temperatures below $T_c$ through fitting to the modified BTK theory [14]. The typical set of one-gap PC spectra measured at different temperatures and used to estimate the temperature dependence of the gap is shown in Fig.4. The minima in the dI/dV(V)-dependences near V=4 mV can be due to the relaxation processes in the contact area. The slowed-down charge equalization between both branches of the quasiparticle spectrum, that is responsible for these minima, is typical for comparatively low-resistance contacts with high-level Andreev current. The contact in Fig.4 belongs to this group.

The corresponding $\Delta(T)$-dependence (Fig.5, solid squares) can be compared, within the experimental error, to the BCS theory (dashed line) excluding the region close to $T_c$. The tail-like deviation of experimental data from the theory in this region is typical of many nonconventional superconductors. One of the reasons for this deviation is the increased width of the superconducting transition in some structural elements, which causes a disagreement between the fitted $T_c^{BCS}$ close to the midpoint of the transition and the temperature of its onset. Crystal structure defects near grain boundaries occurring typically in most multicomponent materials are the main cause for the above smearing. The presence of a thin normal layer at the S-electrode surface can be another influencing factor, though it has no appreciable effect on the critical temperature $T_c^{BCS}$ and Andreev current. This factor is more probable because in the contact discussed, like in many others, the value of the Andreev current is close to the theoretically expected one.

The temperature dependences of the large $\Delta_l(T)$ and the small $\Delta_s(T)$ gaps were found by the same procedure for one of the contacts whose spectra are described adequately in the two-gap approximation (Fig.6). The temperature dependence of the large gap is roughly similar to the BCS theory, while the behavior of the small gap deviates considerably from the theory. The deviation (Fig.6, lower curve) can be attributed to the low stability of the low-energy part of the spectrum near V=0, which shows up as appreciable variations of the dI/dV-amplitude in this range. Such variations are typical of contacts based on magnetic superconductors and are caused by spontaneous or current-induced shifts of the domain wall in the contact area. For this reason it was impossible to fit this small gap-related part of the spectrum to the BTK theory with a good accuracy.

## Conclusions

The spectra of Andreev reflection have been measured in point contacts based on the polycrystalline Fe-based oxipnictide $EuAsFeO_{0.85}F_{0.15}$ having the lowest temperature of the superconducting transition $T_c \approx 11.3$ K among other related materials. We believe that the low $T_c$ is connected to the polyvalency of Eu ions when a part of weakly magnetic trivalent ions traps additional electrons and changes to the bivalent state. In this state Eu ions have very high spin moment $S \sim 7$ $\mu_B$ that result in a strong pair-breaking effect in spin-singlet superconductors.

The fitting of the spectra to the modified BTK theory [14] shows that some spectra can be characterized by a single gap parameter, whereas the two-gap approximation is necessary for most of them. In both cases the gap parameters varied considerably from contact to contact being within 1.3÷3.1 meV ($2\Delta/kT_c$=2.6÷6.4) in the one-gap spectra or within 1.1÷2.3 meV ($2\Delta_s/kT_c$=2.2÷4.7) and 2.5÷5.9 meV ($2\Delta_l/kT_c$=5.1÷11.7) for the small $\Delta_s$ and large $\Delta_l$ gap, respectively, in the two-gap ones. The anomalously high value of the characteristic parameter $2\Delta/kT_c$ obtained for some contacts point to a non-phonon mechanism of pairing in the compound investigated.

We attribute the observed variations of the gap parameters in $EuAsFeO_{0.85}F_{0.15}$ to the anisotropy of the gap function $\Delta(\mathbf{k})$ near the Fermi level. This assumption agrees with some theoretical studies substantiating the existence of the anisotropic s±-symmetry of the order parameter in iron pnictides. The varying from contact to contact intensity of the inter- and intra-band scattering of quasiparticles at impurities and structural defects can be contributory too.

The result obtained in this study imply the absence of zeros or lines of zeros in the $\Delta(\mathbf{k})$-dependence. This excludes the d-wave symmetry of the order parameter in $EuAsFeO_{0.85}F_{0.15}$, which was assumed to exist in some low-$T_c$ pnictides.

The work was partially supported by grants # 09-02-01370 and # 08-08-00709 from Russian Fundamental Research Fund.

# References

[1]. Y. Kamihara, T. Watanable, M. Hirano, H. Hosono, J. Am. Chem. Soc. $\underline{130}$ 3293 (2008).

[2]. X.H. Chen, T. Wu, G. Wu, R.H. Liu, H. Chen, D.F. Fang, Nature $\underline{354}$ 761 (2008).

[3]. E.P. Khlybov, O.E. Omelyanovsky, A. Zaleski, et al., JETP Lett. $\underline{90}$(5) 387 (2009).

[4]. M. Rotter, M. Togel, D. Johrendt, Phys. Rev. Lett., $\underline{101}$ 107006 (2008).

[5]. J.H. Tapp, Z. Tanq, Binq Lv, et al., Phys. Rev. B $\underline{78}$ 0605059(R) (2008).

[6]. F.C. Hsu, J.Y. Luo, K.W. Yeh, et al., Proc. Natl. Acad. Sci. USA $\underline{105}$ 14262 (2008).

[7]. T.X. Chen, Z. Tesanovic, R.H. Liu, et al., Nature $\underline{453}$ 1224 (2008).

[8]. P. Samuely, P. Szabó, Z. Pribulová, et al., Supercond. Sci. Technol. $\underline{22}$ 014003 (2009).

[9]. R.S. Gonnelli, D. Daghero, M. Tortello, et al., Phys. Rev. B $\underline{79}$ 184526 (2009).

[10]. P. Samuely, Z. Pribulová, P. Szabó, et al., Physica C $\underline{469}$ 507 (2009).

[11]. D. Daghero and R.S. Gonnelli, arXiv: 0912.4858, to be published in Supercond. Sci. Technol.

[12]. V.M. Dmitriev, I.E. Kostuleva, E.P. Khlybov, A.J. Zaleski, et al., Low Temp. Phys. $\underline{35}$ 517 (2009).

[13]. M. Yamashita, N. Nakata, Y. Senshu, et al., Phys. Rev. B $\underline{80}$ 220509(R) (2009).

[14]. G.E. Blonder, M. Tinkham, and T.M. Klapwijk, Phys. Rev. B $\underline{27}$ 112 (1983).

[15]. Y. de Wilde, T.M. Klapwijk, A.G.M. Jansen, J. Heil, and P. Wyder, Physica B $\underline{218}$ 165 (1996).

[16]. R.C. Dynes, V. Narayanamurti, and J.P. Garno, Phys. Rev. Lett. $\underline{41}$ 1509 (1978).

[17]. K.A. Yates, L.F. Cohen, Zhi-An Ren, et al., Supercond. Sci. Technol. $\underline{21}$ 092003 (2008).

[18]. P. Szabó, Z. Pribulová, G. Pristáš, et al., Phys. Rev. B $\underline{79}$ 012503 (2009).

[19]. X. Lu, W.K. Park, H.Q. Yuan, et.al. arXiv: 0910.4230.

[20]. G. Deutscher and P. Nozières, Phys. Rev. Lett. $\underline{50}$ 13557 (1994).

[21]. G.E. Blonder and M. Tinkham, Phys. Rev. B $\underline{27}$ 112 (1983).

[22]. D.J. Singh and M.H. Du, Phys. Rev. Lett. $\underline{100}$ 237003 (2008).

[23]. X.G. Luo, M.A. Tanatar, J.-Ph. Reid, et al., Phys. Rev. B $\underline{80}$ 140503(R) (2009).

[24]. F. Malaeb, T. Yoshida, T. Kataoka, et al., J. Phys. Soc. Jpn. $\underline{77}$ 093714 (2008).

[25]. T. Terashima, M. Kimata, N. Kurita, et al., J. Phys. Soc. Jpn. $\underline{79}$ 053702 (2010).

[26]. P. Szabó, P. Samuely, J. Kačmarčik, et al., Phys. Rev. Lett. $\underline{87}$ 137005-1 (2001).

[27]. R. Khasanov, A. Amato, H.H. Klauss, et al., Phys. Rev. Lett. $\underline{102}$ 187005 (2009).

[28]. C. Ren, Z.S. Wang, H.G. Luo, et al., Phys. Rev. Lett. $\underline{101}$ 257006 (2008).

[29]. K. Matano, Z.A. Ren, X.L. Dong, et al., Europhys. Lett. $\underline{83}$ 57001 (2008).

[30]. D.V. Evtushinskiy, D.S. Inosov, V.B. Zabolotnyy, et al., Phys. Rev. B <u>79</u> 054517 (2009).

[31]. Y. Nagai, N. Hayashi, N. Nakai, et al., New J. Phys. <u>10</u> 103026 (2008).

[32]. T.A. Maier, S. Graser, D.J. Scalapino, et al., Phys. Rev. B <u>79</u> 224510 (2009).

[33]. N. Nakai, H. Nakamura, Y. Ota, et al., arXiv: 0909.1195.

[34]. K. Kuroki , H. Usui, S. Onari, et al., Phys. Rev. B <u>79</u> 224511 (2009).

[35]. H. Fukazawa, Y. Yamada, K. Kondo, et al., J. Phys. Soc. Jpn. <u>78</u> 083712 (20